# Epitaxially stabilized iridium spinel oxide without cations in the tetrahedral site

Hiromichi Kuriyama,[1,2] Jobu Matsuno,[2] Seiji Niitaka,[2] Masaya Uchida,[2] Daisuke Hashizume,[2] Aiko Nakao,[2] Kunihisa Sugimoto,[3] Hiroyuki Ohsumi,[4] Masaki Takata,[4] and Hidenori Takagi[1,2]

[1] *Department of Advanced Materials, University of Tokyo, 5-1-5 Kashiwanoha, Kashiwa, Chiba 277-8561, Japan*

[2] *Advanced Science Institute, RIKEN, 2-1 Hirosawa, Wako, Saitama 351-0198, Japan*

[3] *Japan Synchrotron Radiation Research Institute, SPring-8, 1-1-1 Kouto, Sayo-cho, Sayo-gun, Hyogo 679-5198, Japan*

[4] *RIKEN, SPring-8 Center, 1-1-1 Kouto, Sayo-cho, Sayo-gun, Hyogo 679-5148, Japan*

Single-crystalline thin film of an iridium dioxide polymorph $Ir_2O_4$ has been fabricated by the pulsed laser deposition of $Li_xIr_2O_4$ precursor and the subsequent Li-deintercalation using soft chemistry. $Ir_2O_4$ crystallizes in a spinel ($AB_2O_4$) without $A$ cations in the tetrahedral site, which is isostructural to $\lambda$-$MnO_2$. Ir ions form a pyrochlore sublattice, which is known to give rise to a strong geometrical frustration. This Ir spinel was found to be a narrow gap insulator, in remarkable contrast to the metallic ground state of rutile-type $IrO_2$. We argue that an interplay of strong spin-orbit coupling and a Coulomb repulsion gives rise to an insulating ground state as in a layered perovskite $Sr_2IrO_4$.

5*d* transition metal oxides (TMOs) with an odd number of *d*-electrons per transition metal ion are generally believed to be a wide band metal due to the spatially extended character of outermost 5*d* orbitals. Recent discovery of an unconventional Mott insulating state in a layered perovskite $Sr_2IrO_4$, however, has shown that an interplay of a strong spin-orbit coupling (SOC) inherent to heavy 5*d* elements and a Coulomb repulsion $U$ cannot be ignored in 5*d* TMOs.[1-3] In $Sr_2IrO_4$ with $Ir^{4+}(t_{2g}^5)$, SOC as large as ~0.6 eV splits the almost sixfold degenerate $t_{2g}$ bands into effective total angular momentum $J_{eff} = 1/2$ ($J_{1/2}$) and $3/2$ ($J_{3/2}$) states, yielding a half-filled $J_{1/2}$ band. The presence of the narrow half filled band gives rise to an antiferromagnetic Mott insulating state with a modest Coulomb $U \sim 0.5$ eV. Such SOC-induced Mott insulator is now attracting great attention as an exotic state of matter.

If the uniqueness of $Ir^{4+}$ is coupled with a special topology of lattice, such as geometrically frustrated lattice, even more exotic ground state such as correlated topological insulator may be anticipated.[4,5] We put our attention on $AB_2O_4$-type normal spinel compound, where *B* cations in the octahedral site comprise pyrochlore sublattice with corner sharing tetrahedra as shown in Fig. 1. The triangular-lattice-based geometry of pyrochlore lattice is known to produce incompatibility of interactions between the neighboring magnetic *B* ions, called geometrical frustration. To accommodate $Ir^{4+}$ in the spinel *B*-sites, we have to develop a spinel $\square Ir_2O_4$ without *A* cations in the tetrahedral site, isostructural to $\lambda$-$MnO_2$, for the sake of charge neutrality.[6]

We attempted to grow $\lambda$-$MnO_2$-type $Ir_2O_4$ by employing thin film technique and soft chemical method. So far, however, spinel oxide with Ir has not been synthesized yet in the bulk form. The only example of spinel-type Ir oxide known is $ZnIr_2O_4$, which was stabilized by utilizing thin film technique.[7] Our early attempts to grow $Ir_2O_4$ directly on substrates were not successful. Since $\lambda$-$MnO_2$ was reported to be synthesized by a Li-deintercalation from $LiMn_2O_4$ spinel,[6] we took essentially the same approach for $Ir_2O_4$. In this work, by deintercalating Li from epitaxially grown $Li_xIr_2O_4$ films, $Ir_2O_4$ spinel without *A* cations was synthesized. This unique form of $Ir_2O_4$ was found to be an insulator in remarkable contrast to the metallic ground state of rutile-type $IrO_2$.

Epitaxial thin film of $Li_xIr_2O_4$ precursor was grown on $LiNbO_3$(0001) substrate. Films with 70-100 nm thickness were deposited by pulsed laser deposition (PLD) using a KrF excimer laser (λ = 248 nm) at 5 Hz with a fluency of $\sim 3$ J/cm$^2$ on the target surface. An oxygen partial pressure and a deposition temperature of 10 Pa and 923 K, respectively, were used for the optimized thin film growth in this study. A Li-rich target with a composition of Li/Ir ~1 was used to compensate the volatilization of Li during the deposition. The target was prepared by a solid-state reaction. The mixture of $Li_2CO_3$ and $IrO_2$ was calcined in air at 923 K for 12 h. The product was finely ground and sintered in air at 1023 K for 48 h. The obtained target was found to be a mixture of $Li_2IrO_3$ and $IrO_2$. The crystal structures were characterized by various x-ray diffraction (XRD) measurements using Cu $K\alpha$ radiation. The composition of films was checked by x-ray photoelectron spectroscopy (XPS).

In XRD measurement of $2\theta$-$\omega$ scan on obtained films shown in Fig. 2(a), the presence of a single phase of highly oriented Li-Ir-O can be recognized. The full width at half maximum of rocking curve of $\omega$ scan was estimated to be $0.07^\circ - 0.1^\circ$ for the peak at the lowest $2\theta$ angle, indicating a good crystallinity of the Li-Ir-O. The detailed x-ray data indicate that $Li_xIr_2O_4$ spinel is formed as naively expected from the matching with the trigonal $LiNbO_3$ (0001) surface. The Li/Ir composition in Li-Ir-O films was estimated to be $0.3-0.5$ by XPS measurement, excluding the possibility of known Li-Ir-O compounds such as $Li_8IrO_6$ and $Li_2IrO_3$, all with Li rich composition. The largest $d$ value $d_{\max}$= 4.95 Å, evaluated from lowest $2\theta$ peak at $17.91^\circ$, was found to be comparable to the value of $d_{111}$ for spinel. The asymmetric reflections such as (226), (551) and (153) indexed by assuming spinel structure were observed as shown in Fig. 2(d)-(f). The reciprocal mapping around (226) exhibits that the epitaxial relationship between film and substrate is $[11\bar{2}]$ spinel $Li_xIr_2O_4$ $\parallel$ $[11\bar{2}0]$ $LiNbO_3$. This alignment should stabilize the interface with matching of oxygen-arrangement. To further justify the growth of spinel, we checked (004*n*) reflection by fabricating Li-Ir-O thin films on MgO (001) substrate, while (00*l*) reflections cannot be approached as long as films have (111)-orientation. The observation of (226), (551), (153) and (004*n*) peaks are fully consistent with reflection conditions of space group $Fd\bar{3}m$ (No.227) for spinel; *hhl*: $h+l=2n$, *hkl*: $h+k, h+l, k+l=2n$ and 00*l*: $l=4n$. Considering the contrast of ionic charge and ionic radius, though the structural details are not known, it is highly likely that Ir occupies only the octahedral *B* sites and that Li occupies only the tetrahedral *A* sites of spinel.

We found that Li ions can be removed from $Li_xIr_2O_4$ by soft chemical reaction. Li ions were extracted by dipping $Li_xIr_2O_4$ film into 0.02M $I_2$/acetonitrile solution at room temperature. The framework of structure was maintained after the Li-deintercalation which is evidenced by XRD pattern in Fig. 2(a). The small value of full width at half maximum of rocking curve of ω scan of $0.07^\circ - 0.1^\circ$ for (111) reflection indicates that the crystallinity is maintained. Taking look at details, we notice that the peak positions of (*hhh*) reflections shifted to higher $2\theta$s [Fig. 2(b)], indicating that the cell volume was reduced. This reflects the deintercalation of Li. The XPS spectra of the same films clearly indicated the absence of Li 1*s* peak within our resolution. From these results, we concluded that Li ions were entirely removed from the film without any degradation of crystallinity and that $Ir_2O_4$ spinel without *A* cations is formed.

The electrical resistivity and optical conductivity spectrum of $Ir_2O_4$ film on $LiNbO_3$ were measured. We found that $Ir_2O_4$ is a narrow gap insulator in remarkable contrast to the highly conductive metallic state of rutile-type $IrO_2$. As shown in Fig. 3(a), $Ir_2O_4$ exhibits an insulating behavior with a relatively low resistivity of several milliohm centimeter at room temperature. An activation energy of ~0.034 eV was obtained from the resistivity data above 200K. The low resistivity and the low activation energy indicate that the energy gap of $Ir_2O_4$ is likely very small. The narrow energy gap is indeed evidenced by an optical conductivity spectrum $\sigma(\omega)$ measured at room temperature shown in Fig. 3(c). It is clear in Fig. 3(c) that a narrow charge gap is present below a peak ($\alpha$) at 0.5 eV, and that the energy gap should be much smaller than 0.5 eV. These transport and spectroscopic studies indicate that $Ir_2O_4$ is a narrow gap insulator.

The XPS measurement revealed that $Ir_2O_4$ is classified as a $t_{2g}^5$ system with Ir in the octahedral *B* site. The valence-band spectrum is composed of a broad structure ranging from 3 to 9 eV and an intense peak at around 1.5 eV [Fig. 3 (b)]. As observed in $ZnIr_2O_4$,[7] the former originates mainly from O 2*p* states, while the latter is assigned to Ir 5*d* states. Under the octahedral ligand field at the *B* site, the 5*d* states split into $t_{2g}$ and $e_g$ manifolds separated by $\Delta_{LF} \sim 3$ eV. The observed band width of ~3 eV for 5*d* states indicates the absence of the splitting $\Delta_{LF}$ in the valence band: five *d* electrons of $Ir^{4+}$ are accommodated only to $t_{2g}$ states with the low-spin $t_{2g}^5$ configuration. Thus we concluded that $Ir^{4+}$ takes the low-spin $t_{2g}^5$ state in the octahedral *B* site.[8]

The detailed electronic structure of Ir 5d states are observed in optical conductivity $\sigma(\omega)$ [Fig. 3 (c)]. The spectral feature in $\sigma(\omega)$ can be decomposed into two parts below and above $\hbar\omega \simeq 2.5$ eV. The high energy absorption is similar to those observed for $ZnIr_2^{3+}O_4$ with completely filled $t_{2g}$ bands.[7] We assign the absorption spectrum above 3 eV to the interband transition from $t_{2g}$ to empty $e_g$, corresponding to the octahedral crystal field splitting $\Delta_{LF}$. The two peaks, $\alpha$ and $\beta$, at low energies $\hbar\omega = 0.5$ eV $(\alpha)$ and 1.4 eV $(\beta)$ should be assigned to the intraband transition within $t_{2g}$ manifolds, which is absent in $ZnIr_2O_4$.[7]

The presence of two absorption peaks below 2 eV means a complicated and multiple splitting of $t_{2g}$ bands. The two peaks at low energies is strikingly in parallel with those observed in other $Ir^{4+}$ insulator $Sr_2IrO_4$, very likely indicating that the strong SOC splits the $t_{2g}$ manifolds into $J_{1/2}$ bands and $J_{3/2}$ bands as in $Sr_2IrO_4$.[1-3] The electronic structure and the optical transitions adapted from those of $Sr_2IrO_4$ are summarized in the schematics shown in Fig. 3(d). The peak $\alpha$ is assigned to a Mott-Hubbard transition within $J_{1/2}$ bands while the peak $\beta$ corresponds to a transition from filled $J_{3/2}$ bands to the upper Hubbard bands with $J_{1/2}$. The $\alpha$ peak energy of 0.5 eV, which is comparable to that of $Sr_2IrO_4$, indicates that even modest value of *U* can open a Mott gap over a variety of $Ir^{4+}$ oxides.

In summary, we have synthesized $Ir_2O_4$ spinel without *A* cations via $Li_xIr_2O_4$ by utilizing PLD technique and soft chemical method. $Ir_2O_4$ is isostructural to $\lambda$-$MnO_2$, which is composed of $Ir^{4+}$-pyrochlore lattice. It was revealed, by the measurements of optical spectrum and electrical resistivity, that $Ir_2O_4$ is a narrow gap insulator. The insulating state is highly likely a SOC-assisted Mott insulator as in $Sr_2IrO_4$. To further explore the outcome of additional ingredients in this unique $Ir^{4+}$ oxide, in particular the geometrical frustration, the measurement of magnetism is highly desired. The resonant x-ray technique using Ir L edge should be quite powerful probe for this.

This work was financially supported by Grant-in-Aid for Scientific Research (S) (19104008), Grant-in-Aid for Scientific Research on Priority Areas (19052008) and Grant-in Aid for JSPS Fellows.

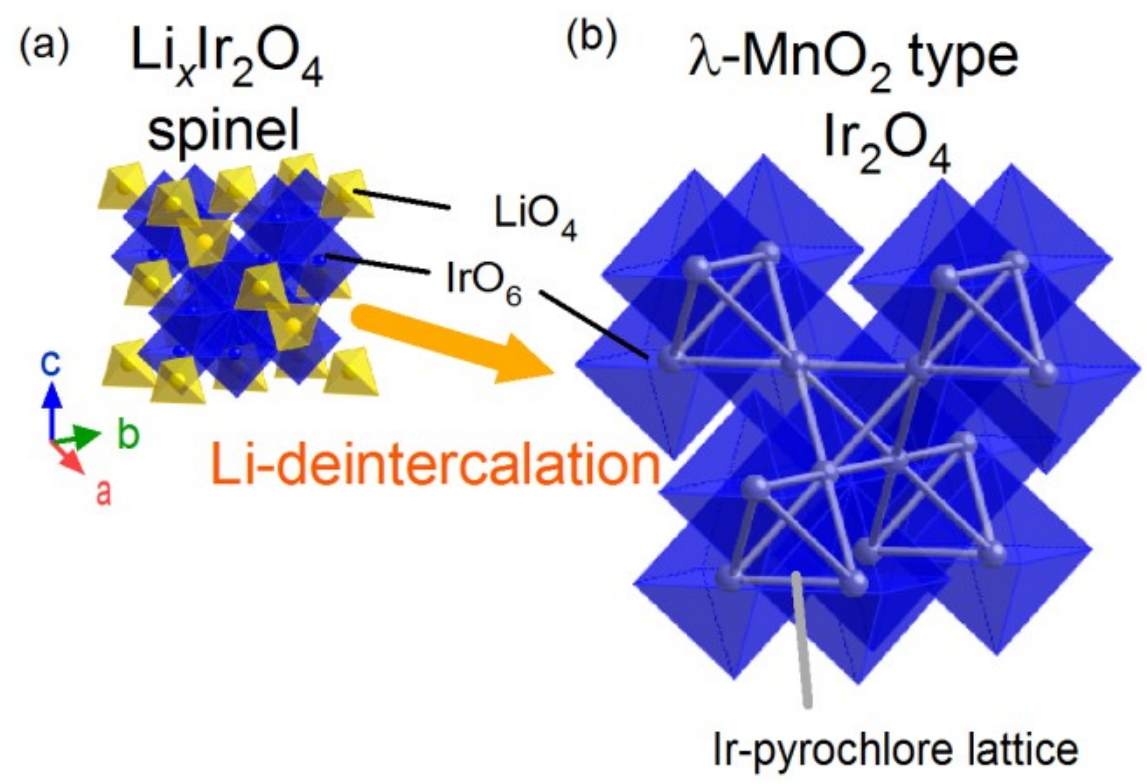

FIG. 1. (Color online) Crystal structures of spinel-type oxide (a) $LiIr_2O_4$ and (b) $Ir_2O_4$. The Ir-pyroclore sublattice is shown by the balls and the sticks in (b). $Ir_2O_4$ represents a Li-deintercalated analog of $LiIr_2O_4$ in this paper.

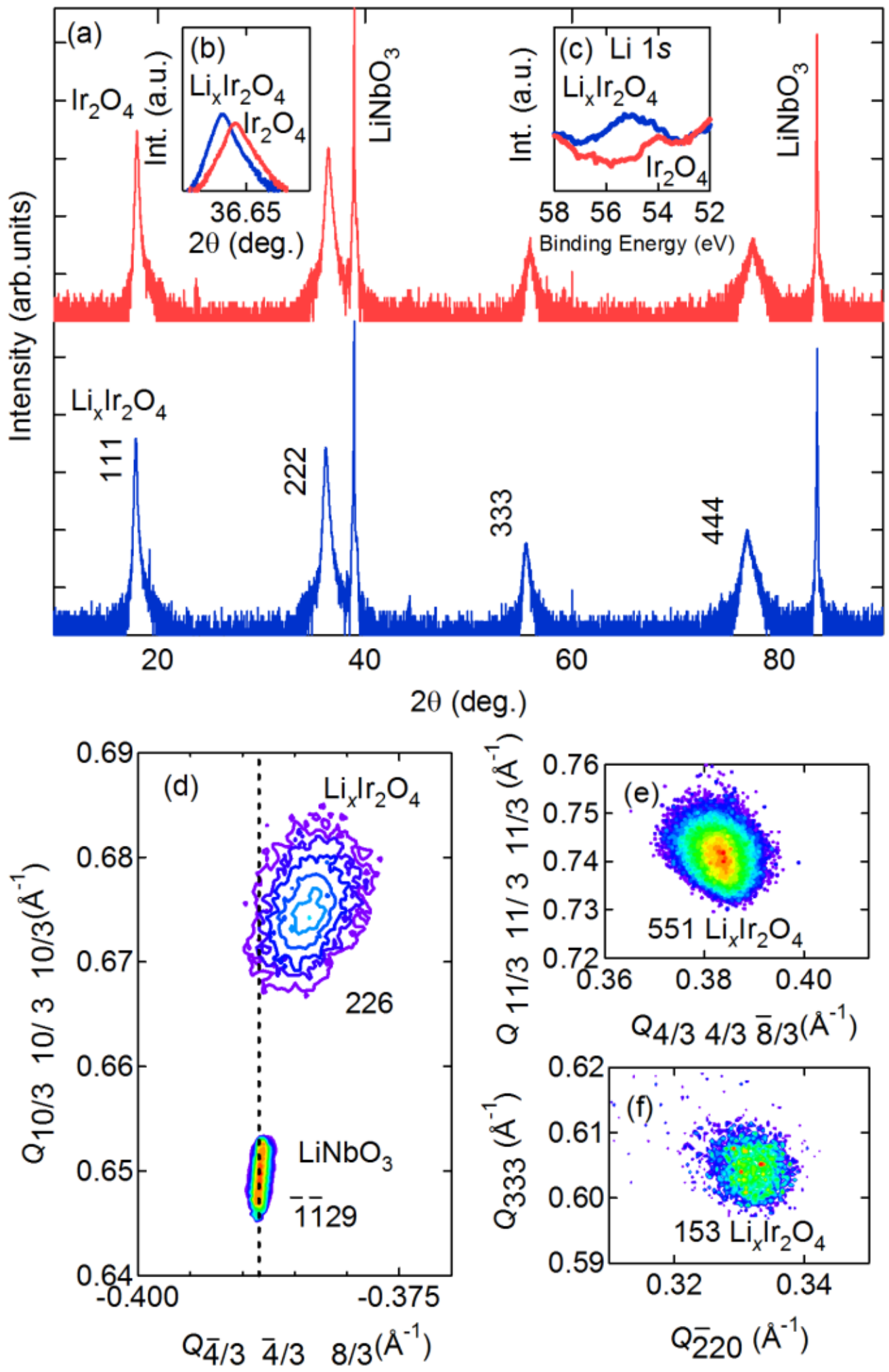

FIG. 2. (Color online) X-ray characterization of $Li_xIr_2O_4$ and $Ir_2O_4$ films on $LiNbO_3$ (0001) substrate. (a) XRD patterns of $2\theta$-$\omega$ scan over a wide range. (b) Enlarged view of (222) reflection. (c) XPS spectra of Li 1$s$ peak. Reciprocal space maps for $Li_xIr_2O_4$ thin film around (d) (226), (e) (551) and (f) (153).

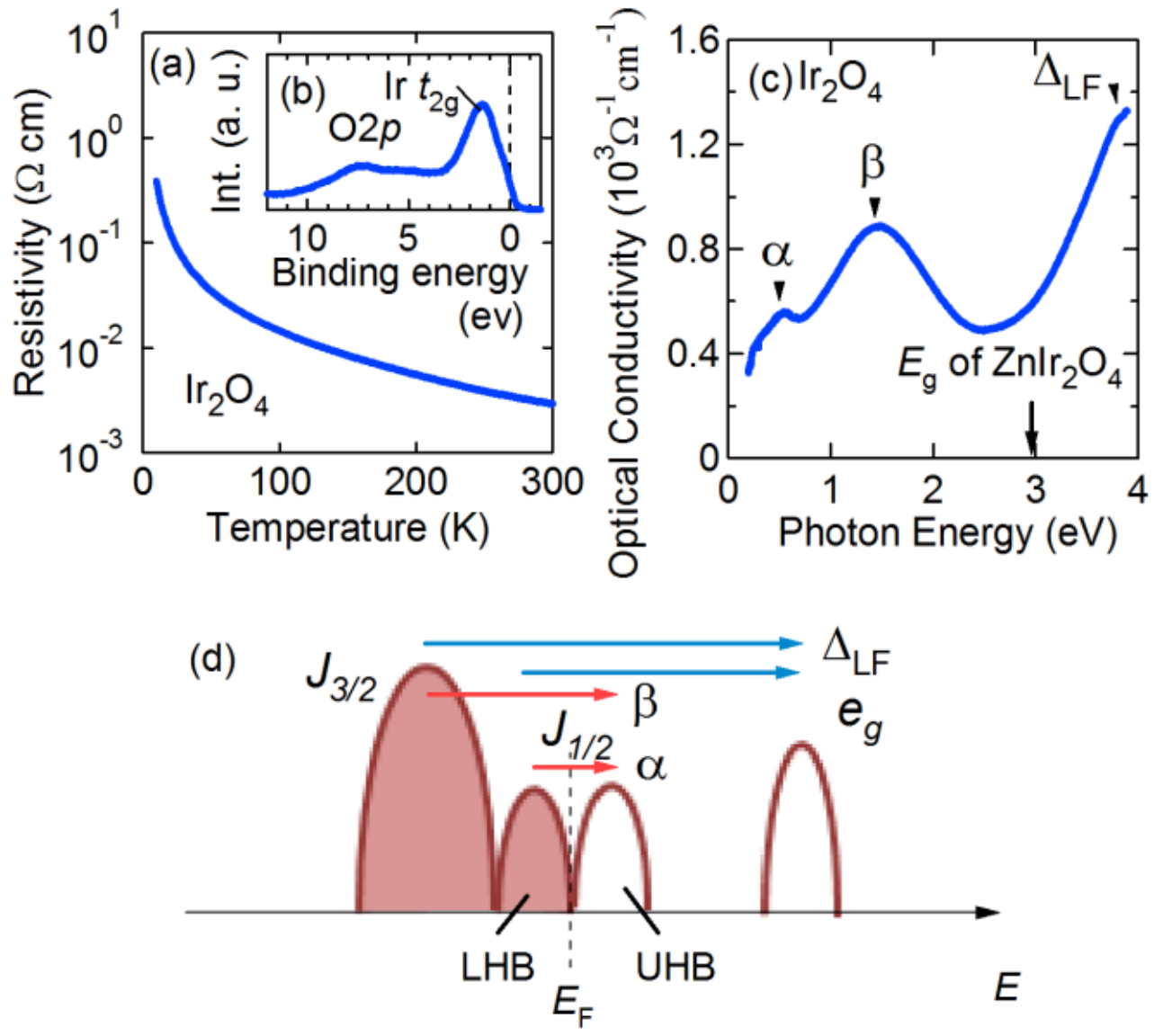

FIG. 3. (Color online) Transport and spectroscopic properties of $Ir_2O_4$ thin film. (a) Temperature dependence of electrical resistivity. (b) Valence-band x-ray photoelectron spectroscopy using Al *K*α excitation source. (c) Optical conductivity measured at room temperature. (c) Schematic view of electronic structure for $Ir_2O_4$ based on Ref. 1. The position of band gap ($E_g \sim 2.97$ eV) for $ZnIr_2^{3+}O_4$ (Ref. 7) is indicated by an arrow in (c) for comparison.